\documentclass[aps, prl, showpacs, twocolumn, amsfonts, amsmath, amssymb, superscriptaddress, floatfix]{revtex4}

\usepackage{graphicx}
\usepackage{dcolumn}
\usepackage{bm}

\newcommand{\ea}{\epsilon_a}
\newcommand{\eb}{\epsilon_b}

\newcommand{\tab}{\stackrel{t_{ab}}{\frown}}

\begin{document}

\title{Localization-Delocalization Transition in the Random Dimer Model}

\author{Jean-Fran\c{c}ois Schaff}
\affiliation{Universit\'e de Nice - Sophia Antipolis, Institut non Lin\'eaire de Nice, CNRS, 1361 route des Lucioles, 06560 Valbonne, France}

\author{Zehra Akdeniz}
\affiliation{Piri Reis University, 34940 Tuzla-Istanbul, Turkey}

\author{Patrizia Vignolo}
\affiliation{Universit\'e de Nice - Sophia Antipolis, Institut non Lin\'eaire de Nice, CNRS, 1361 route des Lucioles, 06560 Valbonne, France}

\begin{abstract}
The random-dimer model is probably the most popular model for a one-dimensional disordered system where correlations are responsible for delocalization of the wave functions. This is the primary model used to justify the insulator-metal transition in conducting polymers and in DNA. However, for such systems, the localization-delocalization regimes have only been observed by deeply modifying the system itself, including the correlation function of the disordered potential. In this article, we propose to use an ultracold atomic mixture to cross the transition simply by externally tuning the interspecies interactions, and without modifying the impurity correlations.
\end{abstract}


\maketitle

In a one-dimensional disordered system, Anderson localization is known to occur at any energy when the disorder is $\delta$ correlated~\cite{Ande58, Ande79}. 
Nevertheless, if one introduces particular short-range correlations, 
delocalization of a 
significant subset of the eigenstates can appear. This happens in the 
random-dimer model (RDM)~\cite{Phil90}, in which the sites of a lattice 
are assigned energies $\epsilon_a$ or $\epsilon_b$ at random, with the 
additional constraint that sites of energy $\epsilon_b$ always appear in 
pairs, or dimers. The same occurs in its dual 
counterpart (DRDM)~\cite{Phil90}, in which lattice sites with energy $\epsilon_b$ 
never appear as neighbors.
In these models, extended states arise from resonant modes of the 
(dual)dimers which
present vanishing backscattering at energy $E_{\rm res}$.
In the thermodynamic limit, the ratio $\sqrt{N}/N$ 
between the number of delocalized states
and the total number of states vanishes, and 
there is no mobility edge separating
extended and localized energy eigenstates.
Nevertheless in finite size systems,
thus in real systems, a localization-delocalization
transition can be induced by driving $E_{\rm res}$ 
inside the spectrum.

This model was proposed to be the possible mechanism which leads to the
insulator-metal transition in a wide class of conducting polymers such as 
polyaniline and heavily doped polyacetylene (see, for instance, 
\cite{Phillips1991a}) and in some biopolymers such as DNA 
\cite{Caetano2005a, Caetano2006a}. The evidence of delocalized electronic states
was experimentally demonstrated in a random-dimer GaAs-AlGaAs superlattice
\cite{Bellani1999a}, while for photons, a RDM dielectric system 
was used \cite{Zhao2007a}. Recently, a RDM setup has been proposed
to demonstrate the delocalization of acoustic waves~\cite{Esmailpour2008a}.
For polymers, semiconductor lattices, photonic crystals and elastic chains, 
the dimer resonant energies cannot be modified  without
changing the sample itself. Thus the localization-delocalization transition
for a (D)RDM chain as a function of the relative position of the resonant 
modes with respect to the band modes cannot easily be studied using 
these physical systems. 

In this article, we propose an experimental procedure 
to realize a DRDM experiment with a one-dimensional (1D) two-component
ultracold atomic mixture in an optical lattice,
and we demonstrate that the localization-delocalization transition
can be explored by tuning the interparticle interactions.

To introduce disorder, a component ($B_d$) has to be classically trapped in the minima of the 
potential \cite{Vign03, Gavish2005a, Paredes2005a}.
For this purpose, one can choose a spin-polarized Fermi component or a strongly 
repulsive hardcore Bose gas. The other component ($B_f$) must be able 
to tunnel through the potential maxima. A single impurity $B_d$ trapped in a 
lattice site causes an energy shift of the effective potential seen by the 
second species $B_f$ with respect to the case where the impurity is absent.
In the following we will focus on a boson-boson mixture, taking recent experiments on the $^{41}$K-$^{87}$Rb mixture at LENS as a guide \cite{Catani2008a}. This mixture has tunable interspecies interactions for both $^{87}$Rb and $^{41}$K in the $|F=1,m_f=1\rangle$ state \cite{Thalhammer2008a}.
The $^{41}$K condensed component plays the role of the ``tunnelling bosons'' $B_f$ and the heavier $^{87}$Rb atoms are the defects $B_d$.

\begin{figure}[t] 
\begin{center}
\begin{minipage}{0.09\linewidth}
(a)
\end{minipage}
\begin{minipage}{0.6\linewidth}
\begin{flushleft}
\includegraphics[width=0.9\linewidth]{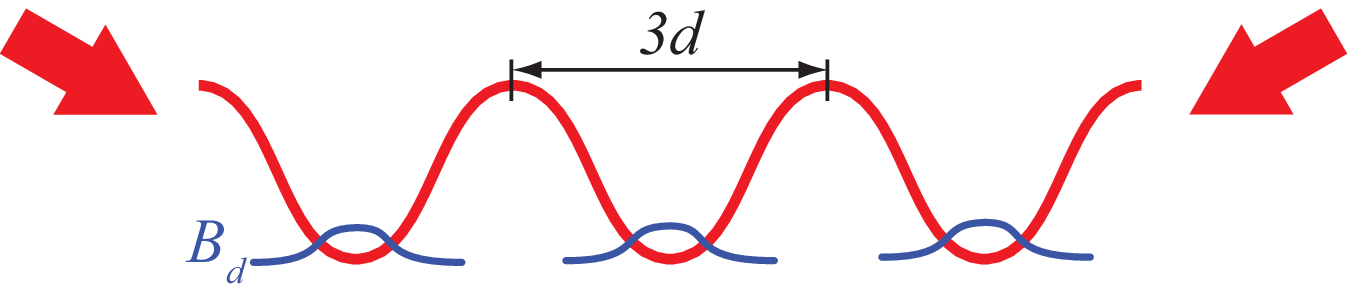}
\end{flushleft}
\end{minipage}\\
\begin{minipage}{0.09\linewidth}
(b)
\end{minipage}
\begin{minipage}{0.6\linewidth}
\begin{flushleft}
\includegraphics[width=0.9\linewidth]{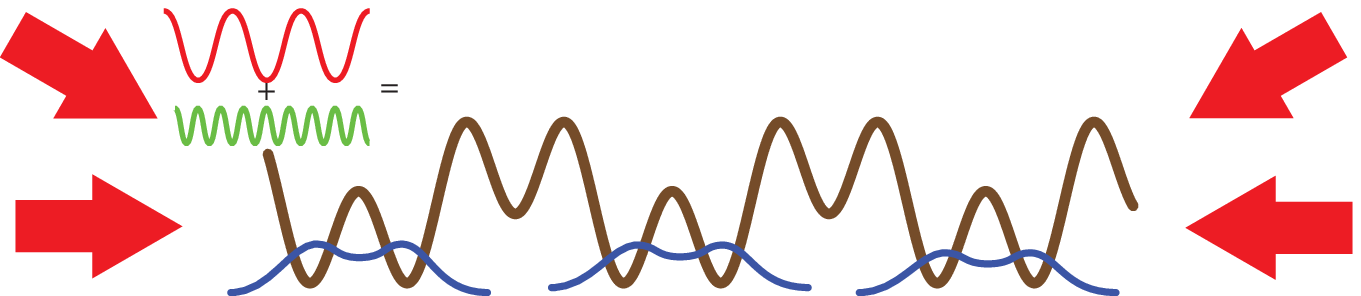}
\end{flushleft}
\end{minipage}\\
\begin{minipage}{0.09\linewidth}
(c)
\end{minipage}
\begin{minipage}{0.6\linewidth}
\begin{flushleft}
\includegraphics[width=0.9\linewidth]{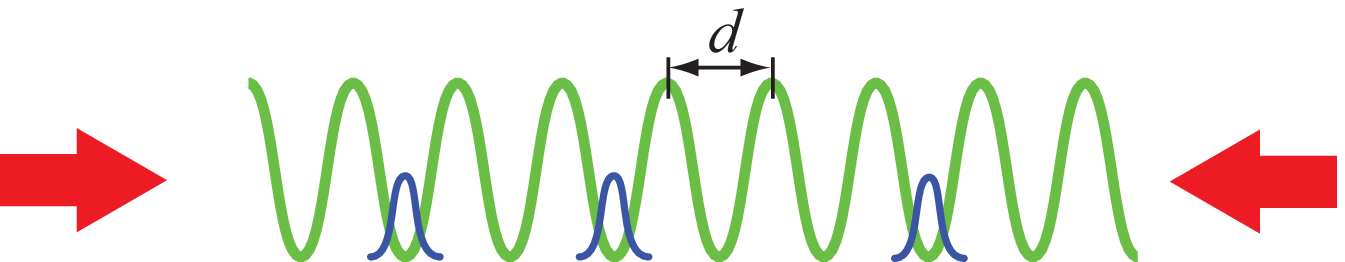}
\end{flushleft}
\end{minipage}
\caption{(Color online) Experimental scheme to generate the correlated disorder. 
(a) Impurities are trapped in a lattice of step $3d$. 
(b) A second lattice of step $d$ is switched on, and the first one turned off. In the final configuration (c), the impurities follow a DRDM distribution.}
\label{fig_lattices}
\end{center}
\end{figure}

To study the effect of correlated impurities $B_d$ on matter-wave transport,
we use the 1D effective tight-binding (TB) Hamiltonian for bosons $B_f$,
\begin{equation}
H_{B_f}=\sum_{i=1}^{n_s}  
E_i |\,i\rangle\langle i\,|+
\sum_{i=1}^{n_s-1}t_i(|\,i\rangle\langle i+1\,|+|\,i+1\rangle\langle i\,|)
\label{Hamiltonian}
\end{equation}
where $n_s$ is the number of sites, $E_i \in \{\epsilon_a, \epsilon_b\}$, and $\epsilon_b$'s never appear as
neighbors. The hopping term $t_i$ can take the values $t_{aa}$ between two sites with energy $\epsilon_a$ or $t_{ab}$ between two sites with different 
energies.
The constraint on sites with energy $\epsilon_b$ fixes the
hopping energies $t_{ab}$ to be distributed as ``dual dimers'' of the form
$\ea \tab \eb \tab \ea$.
To realize such an experiment, one needs to assure that there can be either 
zero or one impurity $B_d$ in each lattice site and that they never appear 
in succession. 
The procedure we propose is schematized in Fig. \ref{fig_lattices}.
(i) First, $N_{B_d}$ atoms of species $B_d$ are trapped in a lattice with a 
step $3d$,
with the condition that no site must be doubly occupied, as in the Tonks gas experiment described in \cite{Paredes2004a}. Such a lattice can be realized using beams of wavelength $\lambda=2d$, tilted by an angle $\alpha={\rm acos}(1/3)$ [see Fig. \ref{fig_lattices}(a)]. For the case $\lambda=800\,$nm and a lattice potential depth $U_{B_d}^{0,in}\simeq 30 E_{B_d}$, $E_{B_d}=4\pi^2\hbar^2/(2m_{B_d}\lambda^2)$ being the recoil energy for a boson of mass $m_{B_d}$, and for an axial confinement of $2\pi\times 60$~Hz \cite{Paredes2004a}, double occupancy can be avoided if $N_{B_d}\le 20$ \cite{Golovach2009}. The number of impurities can be increased by increasing the lattice potential depth or by relaxing the axial confinement.
(ii) Then impurities are forbidden to occupy neighbor sites by adiabatically ramping up the power in a second superimposed lattice of step $d$, and switching off the first one \cite{Foelling2007a}. 
The potential depth $U_{B_d}^0$ of the final lattice $U_{B_d,B_f}(z)=U_{B_d,B_f}^0\sin^2(\pi z/d)$ must be large compared to $E_{B_d}$ to neglect the impurity
mobility during the experiment course (0.5 to 1 s \cite{Billy2008,Roati2008}).
This condition can be fulfilled at $U_{B_d}^0=18E_{B_d}$ in the presence of
attractive interactions with the species $B_f$. 
Differently, the depth $U_{B_f}^0$ for the species $B_f$ must be $\gtrsim$ 2 
$E_{B_f}$, the recoil energy for a $B_f$  boson, to guarantee the validity of the TB description.

The effective Hamiltonian~(\ref{Hamiltonian}) is obtained by a 1D reduction of the system Hamiltonian by introducing the transverse widths $\sigma_{\perp\,B_f,B_d}$ of the condensate and of the impurities wave functions in a cylindrical trap \cite{Salasnich2002}.
Using a TB scheme we introduce the Wannier function $\phi_i(z)$ approximated by the Gaussian function $\phi_i(z)=[\phi_i(0)/(\pi^{1/4}\sigma_{z\,B_f}^{1/2})] \exp[-(z-z_i)^2/(2\sigma_{z\,B_f}^2)]$, where $|\phi_i(0)|^2$ is the number of bosons $B_f$ in the lattice well $i$.
Similarly, the density of impurities is
$n_{B_d}\propto\sum_{i'}\exp[-(z-z_{i'})^2/\sigma_{z,B_d}^2]$.
The determination of the widths $\sigma_{\perp\, B_f,B_d}$ and
$\sigma_{z\,B_f,B_d}$ is carried out variationally~\cite{Salasnich2002,Vign03}.

We can now evaluate the parameters entering the effective
Hamiltonian~(\ref{Hamiltonian}). The site energies are given by
\begin{equation}\begin{split}
E_i=&\int dz\,\tilde\phi_i(z)\left[
-\frac{\hbar^2\nabla^2}{2m_{B_f}}+U_{B_f}(z)\right.\\
+&\left.\frac{1}{2}g
|\phi_i(z)|^2
+g'
n_{B_d}(z)+C_{B_f}\right]\tilde\phi_i(z)
\end{split}
\label{siteenergy}
\end{equation}
where $m_{B_f}$ is the mass of the boson $B_f$, 
$C_{B_f}=\hbar^2/(2m_{B_f}
\sigma_{\perp\,B_f}^2)+\frac{1}{2}m_{B_f}\omega_{\perp\,B_f}^2
\sigma_{\perp\,B_f}^2$,
and $\omega_{\perp\,B_f}$ is the radial frequency of the harmonic 
trapping potential.                                                             $\tilde\phi_i(z)$ are modified Gaussian functions, obtained by imposing 
the condition $\int \tilde\phi_i(z)\tilde\phi_j(z)=\delta_{ij}$ 
\cite{Larson2008a}.       
The parameters $g$ and $g'$ are 
the strengths of the 1D $B_fB_f$ and $B_fB_d$ 
interactions, which are given by
$g=(4\pi\hbar^2 a)/(2\pi m_{B_f}\sigma_{\perp\,B_f}^2)$
and $g'=(2\pi\hbar^2 a')/[\pi m_r(\sigma_{\perp\,B_f}^2+\sigma_{\perp\,B_d}^2)]$,
with $a$, $a'$ the $B_fB_f$ and the $B_fB_d$
scattering lengths and $m_r$ the $B_fB_d$ reduced mass.
The hopping energies $t_i$ are given by
\begin{equation}
t_i=\int dz\,\tilde\phi_i(z)
\left[
-\frac{\hbar^2\nabla^2}{2m_{B_f}}+U_{B_f}(z)\right]
\tilde\phi_{i+1}(z).
\label{hopenergy}
\end{equation}
This completes the determination of the effective 1D Hamiltonian for bosons $B_f$.

Delocalization occurs for energy values near to the resonance energy of a 
single dual dimer embedded in a perfect lattice of site energies $\epsilon_a$
and hopping energies $t_{aa}$.
For such Hamiltonian $H$, the wave function at energy $E$ is $|\varphi\rangle=|k\rangle+G^0\,T\,|k\rangle$,
where $|k\rangle$ is the wave function of the unperturbed periodic Hamiltonian 
$H_0=\sum_{n=-\infty}^{\infty}\epsilon_a|n\rangle\langle n|+
t_{aa}(|n\rangle\langle n+1|+\textrm{c.c.})$,
$G^0$ the unperturbed Green's function $G^0(E)=(E-H_0)^{-1}$, and
$T$ the  matrix $T(E)=H_\textrm{I}(\textbf{1}-G^0H_\textrm{I})^{-1}$, where $H_\textrm{I}$ is a remainder defined as $H_\textrm{I}=H-H_0=(\epsilon_b-\epsilon_a)|0\rangle\langle 0|+(t_{ab}-t_{aa})(|-1\rangle\langle 0|+|0\rangle\langle 1|+\textrm{c.c.})$. Here and below the complex energy $E$ is 
considered in the limit of vanishing positive imaginary part.
Using the renormalization scheme outlined in \cite{Bakhtiari2005a} 
the scattering $T$ matrix in the subspace $\{|-1\rangle,|1\rangle\}$ 
can be written as $T=\tilde H_\textrm{I}(\textbf{1}-G^0\tilde H_\textrm{I})^{-1}$, $\tilde H_{\rm I}$ being the renormalized remainder Hamiltonian. We find 
$\tilde H_{\text{I}} = \alpha \bigl( 
\begin{smallmatrix} 1 & 1\\ 1 & 1
\end{smallmatrix} 
\bigr)$
with $\alpha=t_{ab}^2/(E-\epsilon_b)-t_{aa}^2/(E-\epsilon_a)$.
Thus, the scattering matrix on the subspace $\{|-1\rangle,|1\rangle\}$ is identically
null if $\alpha=0$. This occurs at the resonance energy
\begin{equation}
E_{\rm res}=\dfrac{\epsilon_a t_{ab}^2-\epsilon_bt_{aa}^2}{t_{ab}^2-t_{aa}^2}.
\label{res}
\end{equation}
Eigenstates are delocalized if $E_{\rm res}$ is inside the lowest-energy band
$E(k)=\epsilon_a+2t_{aa}\cos(kd)$, 
namely if 
$|\Delta\epsilon|<2|t^2-1|$, with  
$\Delta \epsilon=(\epsilon_b-\epsilon_a)/t_{aa}$ and $t=t_{ab}/t_{aa}$, 
as found by Dunlap
and collaborators~\cite{Phil90}.
\begin{figure}
\begin{center}
\resizebox{0.85\columnwidth}{!}{\includegraphics{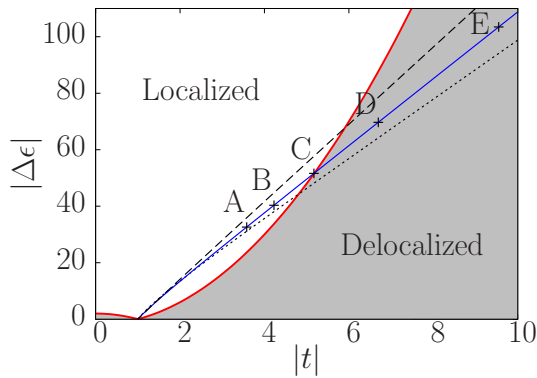}} 
\caption{(Color online) Phase diagram of the DRDM in the plane $(t,\Delta\epsilon)$, with $t=t_{ab}/t_{aa}$ and $\Delta \epsilon=(\epsilon_b-\epsilon_a)/t_{aa}$ and
calculated trajectories for different number of atoms $N_{B_f}$ obtained by varying the interspecies scattering length $a'$. Dashed line, $N_{B_f}=1.6\times10^4$, $a' \in [0,-87\,a_0]$; continuous line, $N_{B_f}=1.3\times10^4$, $a' \in [0,-380\,a_0]$; dotted line, $N_{B_f}=1.0\times10^4$, $a' \in [0,-570\,a_0]$.
With $1.3\times10^4$ atoms, the localization-delocalization transition for the $B_fB_d$ mixture occurs for $a'=-346\,a_0$ (point C). Other points correspond to $a'=-316$ (A), $-331$ (B), $-361$ (D), and $-376\,a_0$ (E), respectively, $a_0$ being the Bohr radius.}
\label{fig-phasedia}
\end{center}
\end{figure}
The corresponding phase diagram in the $(t,\Delta\epsilon)$ plane
is shown in Fig.~\ref{fig-phasedia}.
The central lobe corresponds to the case where the presence of an impurity
is disadvantageous to the hopping of a boson $B_f$. 
In this region delocalization occurs
only for values of $\epsilon_b$ near the center of the band.
The left side of the diagram corresponds to an increase of the hopping 
probability due to the impurities. In this region the energy value $\epsilon_b$
can be in the gap; thus, the disorder strength can be very large,
but delocalization is established just by the value
of $E_{\rm res}$. 
If $E_{\rm res}$ is not an energy value of the spectrum, for a sufficiently long 
lattice, all states are localized. 

For the evaluation of the site and hopping energies, we consider a system
of $N_{B_f}=1.3\times10^4$ $^{41}$K atoms distributed in 200 wells, 10$\%$ of which
are occupied by a $^{87}$Rb atom. We choose the depth $U^0_{B_f}$ equal to
2.5 $E_{B_f}$ and the optical lattice wavelength $\lambda=800$ nm (red detuned for both species).
For linearly polarized beams, 
this fixes the potential depth $U^0_{B_d}$ for the defects to 18 $E_{B_d}$,
and impurity tunneling time of the order of 1 s for points A to E in Fig.~\ref{fig-phasedia}.
The effect of the interactions is enhanced by a tight radial 
confinement $\omega_{\perp\,B_f}/2\pi = 60$~kHz. 
For such a system the phase diagram can be explored just by varying the
$B_fB_d$ scattering length $a'$ (lines in Fig.~\ref{fig-phasedia}).
In the experiments this can be done by exploiting interspecies Feshbach 
resonances \cite{Thalhammer2008a}.
The point (1,0) corresponds to $a'=0$: species $B_f$ do not interact
with impurities; thus, neither site nor hopping energies
are modified by the presence of $B_d$ and the lattice is not disordered.
Higher points of the curve correspond to greater and greater attractive
$B_fB_d$ interactions.

\begin{figure}
\begin{center}
\resizebox{0.95\columnwidth}{!}{\includegraphics{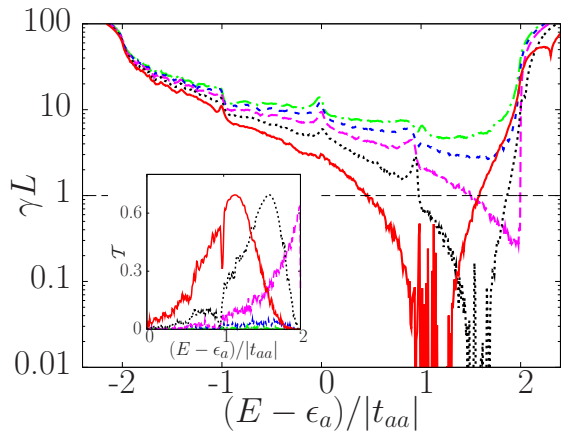}} 
\caption{(Color online)
Lyapunov coefficient $\gamma$ in units of $1/L$, $L=n_sa$ being
the lattice length, as a function of the energy of bosons $B_f$.
The dot-dashed green line corresponds to point A in 
Fig.~\ref{fig-phasedia}, 
the short-dashed blue line
to B, the long-dashed magenta line to C, the dotted black line
to D, and the continuous red line to E. The inset shows the corresponding 
behavior of the transmittivity $\mathcal{T}$.}
\label{fig-lyap}
\end{center}
\end{figure}

We study the spectrum properties across the transition by evaluating
the Lyapunov coefficient
$\gamma(E)$, which is equal to the inverse of the localization
length $\ell(E)$, through the asymptotic relation
\begin{equation}
\gamma(E)=[\ell(E)]^{-1}
=\lim_{n_s\rightarrow\infty}\dfrac{1}{n_sd}\ln\left|
\dfrac{G_{n_s,n_s}(E)}{G_{1,n_s}(E)}\right|,
\label{eq-lyap}
\end{equation}
where $G(E)=(E-H_{B_f})^{-1}$ is the Green's function related to the Hamiltonian
$H_{B_f}$ at energy $E$, and $G_{i,j}(E)=\langle i|G(E)|j \rangle$. 
The matrix element $G_{1,n_s}(E)$ and $G_{n_s,n_s}$
have been computed by exploiting a 
renormalization/decimation scheme \cite{Farchioni1992a}. 
The behaviour of the Lyapunov coefficient through the transition is shown in
Fig.~\ref{fig-lyap}. The different lines, which correspond to the crosses
in Fig.~\ref{fig-phasedia}, show that the localization length is greater
than the system size for points C, D, E. The location of the minima 
corresponds to the position of the resonance energy $E_{\rm res}$, which moves
inside the band for increasing values of $|a'|$. The
nonzero value of $\gamma$ in the delocalization regime 
is due to the finite value of $n_s$ in computing Eq. (\ref{eq-lyap}) 
($n_s$=1000 for the evaluation of $\gamma$).

The nature of the states determines the matter-wave transport properties near equilibrium. These can be evaluated by embedding the whole system
in a perfect lattice of site energies $\epsilon_a$
and hopping energies $t_{aa}$ \cite{Bakhtiari2005a}, 
as previously outlined for the evaluation of the dual-dimer resonance energy. 
The transmission probability $\mathcal{T}=|\tau|^2$, defined as the 
squared modulus of the transmission amplitude,
\begin{equation}
\begin{split}
\tau=&1+G^0_{n_s,1}T_{1,n_s}+G^0_{1,n_s}T_{n_s,1}e^{-2ik(n_s-1)a}\\
+&G^0_{n_s,n_s}T_{n_s,n_s}+G^0_{1,1}T_{1,1},
\label{eq_diff}\end{split}
\end{equation}
for different values of $a'$ is shown in the inset
of Fig.~\ref{fig-lyap}. For $a'=-346\,a_0$ the resonance fits in the 
band-edge, and the corresponding transmission peak arises. For
$a'=-361\,a_0$ and $a'=-376\,a_0$ the peak moves toward the center of the 
band in agreement with the position of the minimum value of $\gamma$. The width of the peak decreases by increasing the system size, as the percentage of the delocalized states scales as $\sqrt{n_s}/n_s$.

Since the condensate energy corresponds to the lowest allowed energy
(quasimomentum $k = 0$), the region nearby the resonance could
be explored by preparing Bloch states with initial
quasimomentum $k \neq 0$ by introducing a constant frequency shift 
between the two waves generating the lattice \cite{Peik1997a}.
We expect that, in the localization regime, for any 
$k\in[-\pi/d,\pi/d]$, the whole condensate stays 
at rest in the reference frame of the moving lattice, while, 
in the delocalization regime, at $k=k(E_{\rm res})$,
the bulk of the condensate stays at rest in the laboratory reference frame.

\begin{figure}
\begin{center}
\resizebox{0.95\columnwidth}{!}{\includegraphics{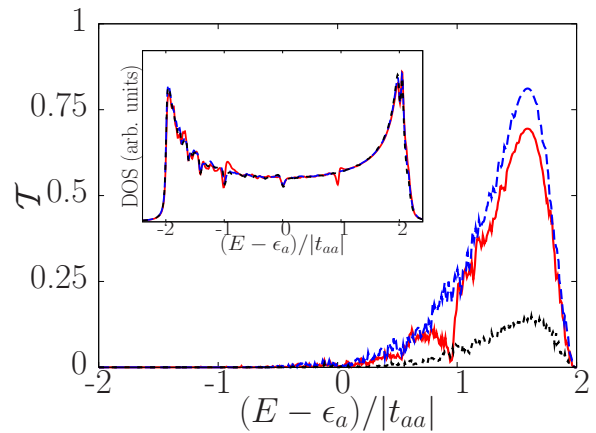}} 
\caption{(Color online) 
Transmission coefficient $\mathcal{T}$ for $N = 1.3 \times 10^4$ and $a' = -361\,a_0$, and different disorder 
patterns: a genuine DRDM lattice (dashed blue line),
a DRDM lattice generated with the procedure illustrated in Fig.~\ref{fig_lattices} (continuous red line), and an uncorrelated lattice
(dotted black line). The inset shows the corresponding density of states.}
\label{compairison}
\end{center}
\end{figure}

Both the Lyapunov coefficient and the transmittivity show some
small peaks (sinkings). These structures are due to an underlying
order present in the procedure illustrated in Fig.~\ref{fig_lattices}. 
In fact, even if our proposition allows the distance between two
subsequent impurities to be equal to any integer $> 1$, still
every three sites is definitly without an impurity.
The evidence that this underlying order does not affect the
DRDM physical effect is shown in Fig.~\ref{compairison}.
The transmittivity peak for the DRDM pattern proposed in this article 
(Fig.~\ref{fig_lattices}) is in correspondence with the
transmittivity peak for a genuine DRDM.
Such a disorder pattern could be generated by using a dipolar gas
for which repulsive interactions may avoid next-neighboring
occupation \cite{Lahaye2009a}. However, at the moment of writing, 
no dipolar gases have yet been cooled down to the degenerate 
regime in mixtures.

For completeness of our analysis, we compare the two DRDM models 
with a lattice where the position of impurities $\epsilon_b$ are uncorrelated \cite{Krutisky2008a}.
In this case the transmittivity drops (Fig.~\ref{compairison}) and becomes vanishing for longer chains. The residual peak is a signature of the presence of a few dual dimers, and it would be washed out in an ordinary disorder model with uncorrelated on-site 
and hopping energies.
It is worth pointing out that the impurity distribution deeply modifies the nature 
of the states, but not the spectrum itself at this low impurity concentration.
Indeed, the density of states (DOS) is essentially the same for the three cases as shown in the inset of Fig.~\ref{compairison}. When the percentage of impurities is increased, the underlying periodicity, which is different in the three models, leads to fragmentation of the DOS in three, two, or one band.
The DOS, $\mathcal{N}(E)$, has been evaluated by using the Kirkman-Pendry relation
$\mathcal{N}(E)=\frac{1}{\pi}{\rm Im}\{[\partial 
\ln G_{1,n_s}(E)]/(\partial E)\}$ \cite{Kirkman1984}.


In conclusion, in this work we show that, at fixed
correlation function among defects, localization-delocalization
can be induced by varying the impurity cross section.
In an ultracold boson-boson mixture, where one component
plays the role of correlated impurities, the rule being that no 
next-neighboring impurities are allowed (DRDM), this can be realized
by driving the interspecies interaction by means of Feshbach resonances.
This is a unique opportunity compared to other physical domains where this class of disorder was previously identified as the possible explanation of the mechanism causing the amazing conducting properties of disordered 1D systems, such as conjugated polymers or biopolymers.

\acknowledgments{
This work was supported by the CNRS and the TUBITAK (exchange of researchers, grant No. 22441) and by the \textit{F\'ed\'eration de Recherches Wolfgang D\"oblin} (CNRS FR 2800). Z.A. acknowledges the support received from the European Science Foundation (ESF) for the activity entitled `Quantum Degenerate Dilute Systems'. P.V. is indebted to F. Mortessagne and P. Sebbah for many useful discussions.}


\end{document}